\documentclass[twoside,twocolumn,9pt]{article}
\usepackage{extsizes}
\usepackage[super,sort&compress,comma]{natbib} 
\usepackage[version=3]{mhchem}
\usepackage[left=1.5cm, right=1.5cm, top=1.785cm, bottom=2.0cm]{geometry}
\usepackage{balance}
\usepackage{mathptmx}
\usepackage{sectsty}
\usepackage{comment}
\usepackage{graphicx} 
\graphicspath{{figures/}} 
\usepackage{lastpage}
\usepackage[format=plain,justification=justified,singlelinecheck=false,font={stretch=1.125,small,sf},labelfont=bf,labelsep=space]{caption}
\usepackage{float}
\usepackage{fancyhdr}
\usepackage{fnpos}
\usepackage[english]{babel}
\addto{\captionsenglish}{%
  
}
\usepackage{array}
\usepackage{droidsans}
\usepackage{charter}
\usepackage[T1]{fontenc}
\usepackage[usenames,dvipsnames]{xcolor}
\usepackage{setspace}
\usepackage[compact]{titlesec}
\usepackage{hyperref}

\usepackage{epstopdf}
\usepackage{physics}
\definecolor{cream}{RGB}{222,217,201}
\usepackage{xspace}
\usepackage{siunitx}

\begin{document}

\pagestyle{fancy}
\thispagestyle{plain}
\fancypagestyle{plain}{
\renewcommand{\headrulewidth}{0pt}
}

\makeFNbottom
\makeatletter
\renewcommand\LARGE{\@setfontsize\LARGE{15pt}{17}}
\renewcommand\Large{\@setfontsize\Large{12pt}{14}}
\renewcommand\large{\@setfontsize\large{10pt}{12}}
\renewcommand\footnotesize{\@setfontsize\footnotesize{7pt}{10}}
\makeatother

\renewcommand{\thefootnote}{\fnsymbol{footnote}}
\renewcommand\footnoterule{\vspace*{1pt}%
\color{cream}\hrule width 3.5in height 0.4pt \color{black}\vspace*{5pt}} 
\setcounter{secnumdepth}{5}

\makeatletter 
\renewcommand\@biblabel[1]{#1}            
\renewcommand\@makefntext[1]%
{\noindent\makebox[0pt][r]{\@thefnmark\,}#1}
\makeatother 
\renewcommand{\figurename}{\small{Fig.}~}
\sectionfont{\sffamily\Large}
\subsectionfont{\normalsize}
\subsubsectionfont{\bf}
\setstretch{1.125} 
\setlength{\skip\footins}{0.8cm}
\setlength{\footnotesep}{0.25cm}
\setlength{\jot}{10pt}
\titlespacing*{\section}{0pt}{4pt}{4pt}
\titlespacing*{\subsection}{0pt}{15pt}{1pt}

\fancyfoot{}
\fancyfoot[LO,RE]{\vspace{-7.1pt}\includegraphics[height=9pt]{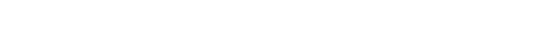}}
\fancyfoot[CO]{\vspace{-7.1pt}\hspace{13.2cm}\includegraphics{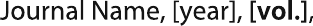}}
\fancyfoot[CE]{\vspace{-7.2pt}\hspace{-14.2cm}\includegraphics{head_foot/RF}}
\fancyfoot[RO]{\footnotesize{\sffamily{1--\pageref{LastPage} ~\textbar  \hspace{2pt}\thepage}}}
\fancyfoot[LE]{\footnotesize{\sffamily{\thepage~\textbar\hspace{3.45cm} 1--\pageref{LastPage}}}}
\fancyhead{}
\renewcommand{\headrulewidth}{0pt} 
\renewcommand{\footrulewidth}{0pt}
\setlength{\arrayrulewidth}{1pt}
\setlength{\columnsep}{6.5mm}
\setlength\bibsep{1pt}

\makeatletter 
\newlength{\figrulesep} 
\setlength{\figrulesep}{0.5\textfloatsep} 

\newcommand{\topfigrule}{\vspace*{-1pt}%
\noindent{\color{cream}\rule[-\figrulesep]{\columnwidth}{1.5pt}} }

\newcommand{\botfigrule}{\vspace*{-2pt}%
\noindent{\color{cream}\rule[\figrulesep]{\columnwidth}{1.5pt}} }

\newcommand{\dblfigrule}{\vspace*{-1pt}%
\noindent{\color{cream}\rule[-\figrulesep]{\textwidth}{1.5pt}} }

\makeatother

\twocolumn[
  \begin{@twocolumnfalse}
{\includegraphics[height=30pt]{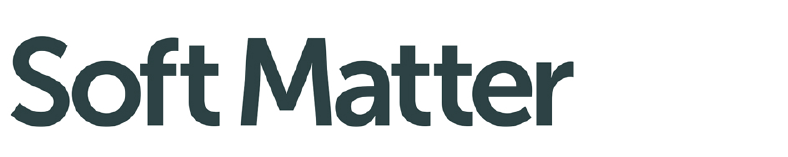}\hfill\raisebox{0pt}[0pt][0pt]{\includegraphics[height=55pt]{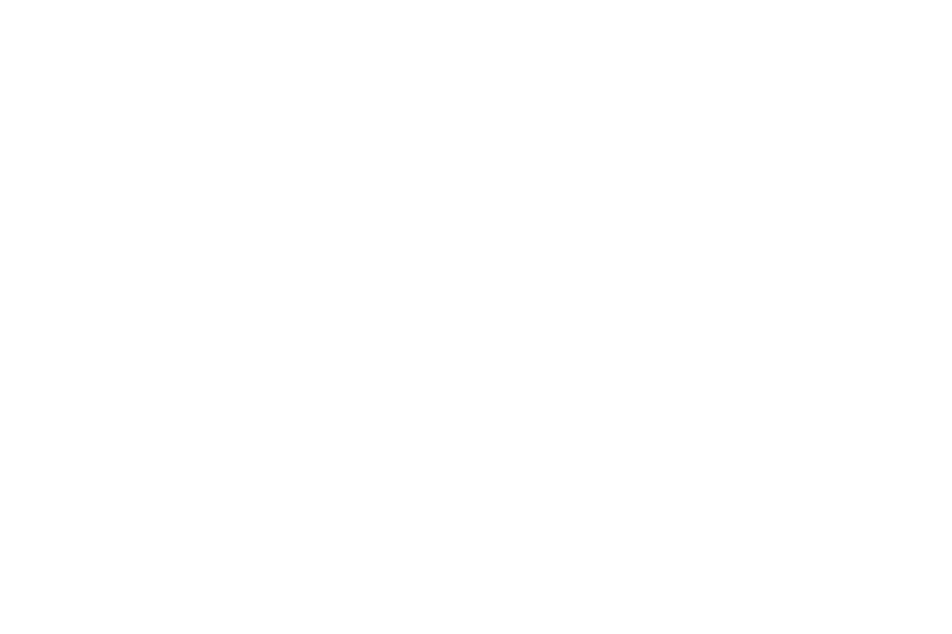}}\\[1ex]
\includegraphics[width=18.5cm]{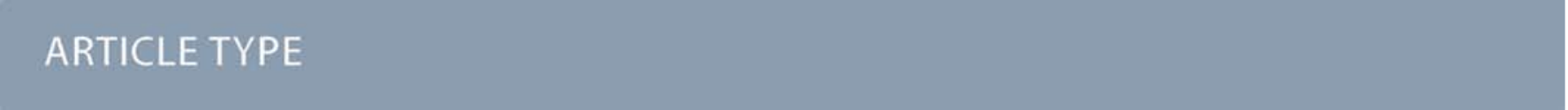}}\par
\vspace{1em}
\sffamily
\begin{tabular}{m{4.5cm} p{13.5cm} }

\includegraphics{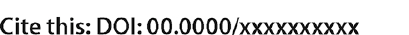} & 
\noindent\LARGE{\textbf{Effect of confinement on the mechanics of a swelling hydrogel bead$^\dag$}}\\
\vspace{0.3cm} & \vspace{0.3cm} \\

 & \noindent\large{Chaitanya Joshi,\textit{$^{a\ddag}$} Mathew Q. Giso,\textit{$^a$} Jean-Fran\c{c}ois Louf,\textit{$^{b}$} Sujit S. Datta\textit{$^{c}$} and Timothy J. Atherton\textit{$^{a\ast}$}} \\

\includegraphics{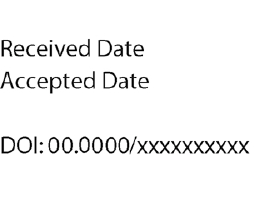} & \noindent\normalsize{We recast the problem of hydrogel swelling under physical constraints as an energy optimization problem. We apply this approach to compute equilibrium shapes of hydrogel spheres confined within a jammed matrix of rigid beads, and interpret the results to determine how confinement modifies the mechanics of swollen hydrogels. In contrast to the unconfined case, we find a spatial separation of strains within the bulk of the hydrogel as strain becomes localized to an outer region. We also explore the contact mechanics of the gel, finding a transition from Hertzian behavior to non-Hertzian behavior as a function of swelling. Our model, implemented in the \textit{Morpho} shape optimization environment, can be applied in any dimension, readily adapted to diverse swelling scenarios and extended to use other energies in conjunction.}

\end{tabular}

 \end{@twocolumnfalse} \vspace{0.6cm}

  ]

\renewcommand*\rmdefault{bch}\normalfont\upshape
\rmfamily
\section*{}
\vspace{-1cm}


\footnotetext{\textit{$^{a}$~Department of Physics and Astronomy, Tufts University, Medford, Massachusetts 02155, USA; E-mail: timothy.atherton@tufts.edu}}
\footnotetext{\textit{$^{b}$~Department of Chemical Engineering, Auburn University, Auburn, AL 36849, USA}}
\footnotetext{\textit{$^{c}$~Department of Chemical and Biological Engineering, Princeton University, Princeton, NJ 08544, USA}}
\footnotetext{\dag~Electronic Supplementary Information (ESI) available: [details of any supplementary information available should be included here]. See DOI: 10.1039/cXsm00000x/}



\section{Introduction}

Hydrogels are polymer networks that have an incredible capacity to absorb water while remaining intact \cite{Bertrand2016}. They are suitable for a variety of practical applications such as hygienic products and contact lenses and other areas \cite{Ahmed2015}. Their similarity to biological tissues have made them a promising material candidate for biomedical and bio-interface devices \cite{Daly2020,Yuk2022} and drug delivery applications \cite{Li2016}. Additionally, hydrogels are used as soil conditioners to improve water retention and other desirable agricultural properties \cite{Azzam1983,Wei2013}. The success of these improvements is known to depend on the size of the soil particles they are embedded in \cite{Rizwan2021}, but the mechanisms by which confinement alters the behavior of the gel are challenging to study directly. 

\begin{figure*}
    \centering
    \includegraphics{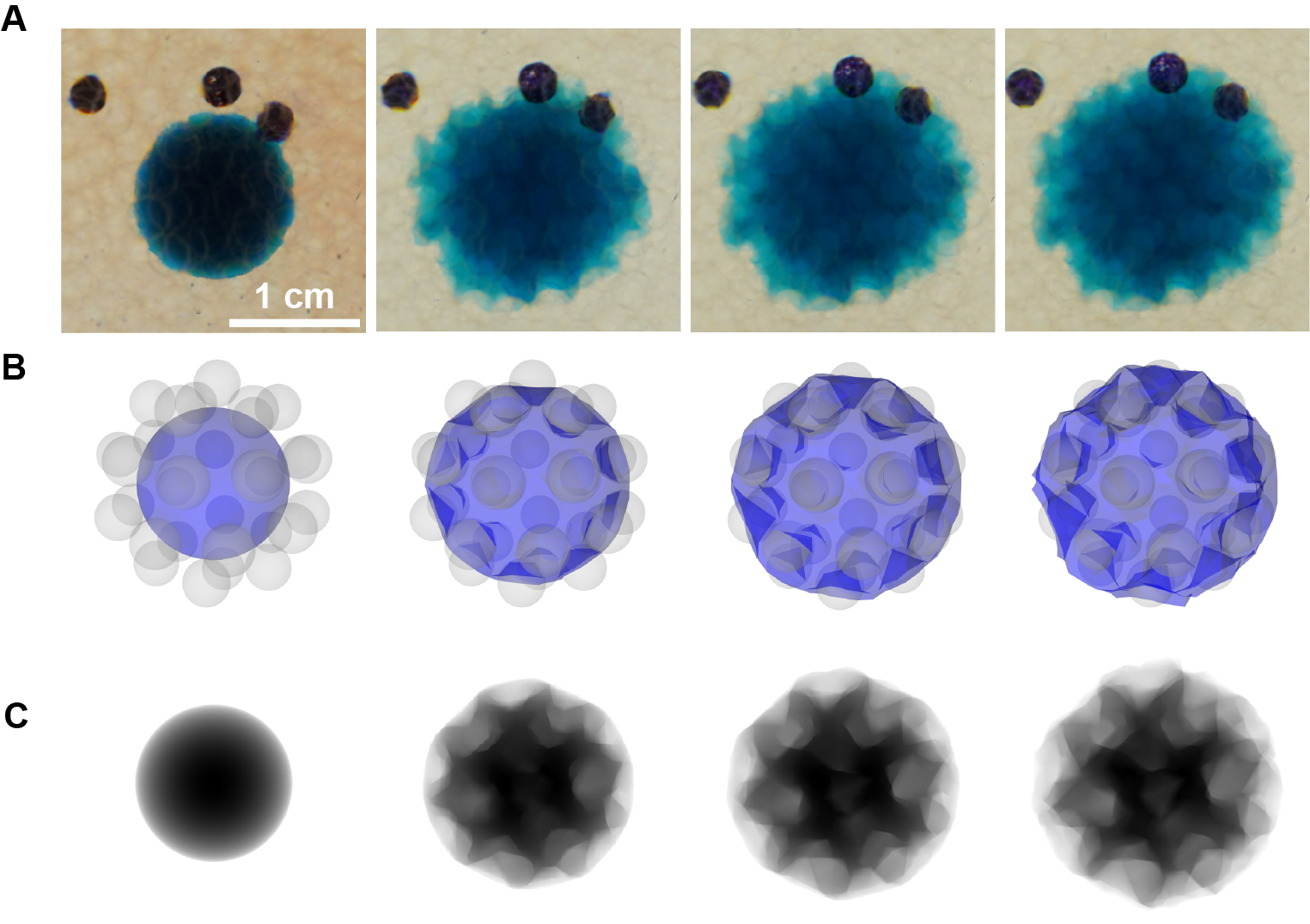}
    \caption{Hydrogel swelling under 3D confinement. \textbf{A} Snapshots of an initially-spherical hydrogel (blue) embedded within a granular medium composed of glass beads (hazy transparent circles) packed within a transparent acrylic chamber. As the hydrogel swells, it deforms strongly due to confinement. Black circles show dyed beads used as tracers to quantify any deformations of the granular packing. The images are taken following the same experimental protocol as in Ref.~\cite{Louf2021}. \textbf{B} Simulation snapshots of a swelling hydrogel surrounded by stationary hard-sphere beads  (grey spheres) at various stages of energy minimization. \textbf{C} Corresponding simulated images obtained by rasterizing the configurations in subfigure B and projecting the the viewing plane. }
    \label{fig:experiment}
\end{figure*}

A recent experimental work on swelling of hydrogel confined in a granular medium \cite{Louf2021} characterized the 3D swelling of a hydrogel sphere surrounded by transparent beads under confining pressure. If the confining pressure is weak, the hydrogel sphere tends to rearrange the surrounding matrix as it swells; for strong confinement the sphere deforms and tends to fill the interstices of the bead packing as is shown in Fig. \ref{fig:experiment}A. The overall degree of swelling was found to be well described by a model that balanced swelling pressures and contact forces, making some assumptions about the distribution of deformation. While the deformation state and the swelling ratio of the hydrogel sphere can be measured, the internal stress and strain distributions are not directly accessible in the experiments. 

Modeling provides a complementary approach, but because hydrogels are highly deformable, they change shape considerably when swollen or indented. Due to the difficulty of capturing these dramatic deformations, modeling of hydrogels has often been restricted to simple geometries \cite{Blanco2013}. Incorporation of constraints that arise naturally in applications, such as the surrounding matrix of soil particles, is also a challenge for modeling \cite{Kang2010}. 

In this paper, we will build a systematic understanding of how confinement modifies the behavior of swollen hydrogels. To do so, we recast hydrogel expansion as an optimization problem, discretize the gel using finite elements and solve the resulting problem to identify thermodynamic equilibrium states. Other finite element models of hydrogel swelling have enabled researchers to access stress and strain distributions \cite{Rognes2009a,Blanco2013}. Variational approaches have been previously used to compute thermodynamic hydrogel profiles under certain class of constraints \cite{Kang2010,Zalachas2013}. The mechanics of contact for constrained hydrogels have been also explored analytically for prototypical geometries \cite{Zheng2019}. However, there is presently no general purpose finite element scheme to solve for equilibrium hydrogel shapes with realistic constraints and/or additional energetic influence, such as surface tension, applied fields, etc. Here, we will use \emph{Morpho}, a programmable environment for shape optimization \cite{DeBenedictis2016a,Joshi2022b} to construct and solve the model. 

The rest of the paper is organized as follows: In section 2, we review the Flory-Rehner theory of hydrogel swelling, formulate the equilibrium problem as a shape-driven energy optimization problem and describe the computational method. In section 3, we describe the resulting simulations for hydrogel spheres swelling in the presence of jammed beads and examine results. Finally, in section 4, we discuss other applications of our method and possible extensions to it. 

\section{Model}

Theoretical modeling of hydrogel configurations in the literature often uses a pressure balancing approach, whereby the mixing pressure and entropic elasticity compete to determine the degree of swelling. 
Here, we wish to instead pose the problem as optimizing a free energy to identify stationary states. In order to do so, we review the Flory-Rehner theory of hydrogel swelling and present it in a form amenable to discretization. The theory constructed reduces to conventional presentations of pressure balance as shown in the Appendix. 

\subsection{Theory}

Consider a polymer hydrogel in a solvent at a fixed temperature $T$ with an internal mesh of permanently crosslinked polymer chains. We note that hydrogels with dynamic or transient crosslinks are also of interest because they give rise to new relaxation dynamics and viscoelastic effects, but are not treated here. Let the number of polymers be $N_\text{p}$ and the number of solvent molecules inside the hydrogel be $N_\text{s}$. 
Let the volume occupied by one monomer/molecule be $\nu_\text{s}$, in the sense of Flory's lattice model \cite{fernandez-nieves_2011,Rognes2009a}. Since $\nu_\text{s}$ is fixed, 
along with the number of polymers $N_\text{p}$ in the hydrogel, the only free parameter during the hydrogel swelling is the number of solvent molecules $N_\text{s}$. Since the swelling process occurs at a fixed total volume (hydrogel plus the external solvent) and temperature, the usual Helmholtz free energy for the mixing of a polymer with a solvent can be used:
\begin{equation}
\label{eq:flory-huggins}
     \Delta F_{\text{mix}} = k_{\text{B}} T \left[ N_\text{p} \ln \phi + N_\text{s} \ln (1-\phi) + \chi N_\text{s} \phi \right]
\end{equation}
where $k_{\text{B}}$ is the Boltzmann constant and $\phi = x N_\text{p} / (N_\text{s} + x N_\text{p})$ is the volume fraction of the polymer molecules, with $x$ being the number of units per polymer. $\chi$ is the Flory-Huggins mixing parameter \cite{fernandez-nieves_2011}. 
Since we are considering highly swollen hydrogels, we can assume that $N_\text{p} \ll N_\text{s}$, thus simplifying the free energy as follows:
\begin{equation}
\label{eq:flory-huggins-simplified}
     \Delta F_{\text{mix}} = N_\text{s} k_{\text{B}} T \left[ \ln (1-\phi) + \chi \phi \right]
\end{equation}
As we noted above, this formalism can be connected to the osmotic pressure formalism by noting that the swelling process also occurs at a constant pressure \cite{fernandez-nieves_2011}, and thus we can equate the Helmholtz free energy to the Gibbs free energy: $\Delta F_\text{mix} = \Delta G_\text{mix}$. The mixing process alters the chemical potential $\mu$ of the solvent, resulting in an osmotic pressure, which can be derived from the Gibbs free energy:
\begin{equation}
    \Pi_\text{mix} = -\frac{N_{\text{A}} \Delta \mu}{\nu_\text{s}} = -\frac{N_{\text{A}}}{\nu_\text{s}} \pdv{\Delta G_\text{mix}}{N_\text{s}} = -\frac{N_{\text{A}}}{\nu_\text{s}} \pdv{\Delta F_\text{mix}}{N_\text{s}}
\end{equation}
with $N_\text{A}$ being the Avogadro’s number.
Similarly, the free energy associated with elasticity can be written like so:
\begin{equation}
    \Delta G_\text{el} = \Delta F_\text{el} = \frac{3 k_{\text{B}} T N_\text{c}}{2} [\alpha^2 - 1 - \ln \alpha]\
\end{equation}
where $N_\text{c}$ is the number of polymer chains, where a chain is defined as the polymer between two cross-link points\cite{fernandez-nieves_2011}, and $\alpha = (V/V_0)^{1/3} = (\phi_0 / \phi)^{1/3}$ is the linear swelling ratio, with $V_0$ and $\phi_0$ being a reference volume and fraction \cite{Quesada-Perez2011}. 
The change in the free energy, under a separability approximation, can we written as 
\begin{equation}
    \Delta F = \Delta F_{\text{mix}} + \Delta F_{\text{el}} 
    \label{eq:hg}
\end{equation}
Equilibrium is defined by the extremization of this free energy, which is equivalent to the balance of osmotic pressures, $\Pi = \pdv*{\Delta F}{N_s} = 0$. Note that due to the direct relationship between $N_\text{s}$ and $\phi$, the free energy can be written solely in terms of $\phi$, and thus, we can cast hydrogel swelling as a free energy minimization problem with respect to $\phi$.

\subsection{Finite Element Modeling}

We now consider a hydrogel where the volume fraction of the polymer can vary over space, defining $\vb{x}$ as the 3D spatial coordinate. Hence, we work with a free energy \emph{density} $\Delta f_{\text{mix}} (\vb{x})$, which is now a function of a spatially varying field $\phi(\vb{x})$. If this space is discretized using simplicial elements---in this work we use tetrahedra in 3D, but the theory is dimensionally independent and readily applicable to other kinds of elements---it is useful to consider the expression \eqref{eq:flory-huggins} for a single element. We will work in the deformed frame of reference \cite{Rognes2009a} as this is the most natural frame to express interpenetrability constraints as desired for the application. Hence, the energy density locally at a point $\vb{x}$ \emph{in the deformed frame of reference} will be Eq.~\eqref{eq:flory-huggins} evaluated at $\vb{x}$ divided by the volume of the element. Since this volume would also be given by $\nu_\text{s} (x N_\text{p}  + N_\text{s})$, we have,

\begin{align}
    \Delta f_{\text{mix}} &= \frac{N_\text{s}}{\nu_\text{s}(x N_\text{p} + N_\text{s})} k_{\text{B}} T \left[ \ln (1-\phi) + \chi \phi \right] \\
    &= \frac{(1-\phi)}{\nu_\text{s}} k_{\text{B}} T \left[ \ln (1-\phi) + \chi \phi \right] 
\end{align}
This can be expressed in terms of an `effective diameter' of the solvent molecule $d$ such that $d^3 = \nu_\text{s} / N_\text{A}$ \cite{Louf2021}. In terms of $d$, this reduces to,
\begin{equation}
    \Delta f_{\text{mix}} = \frac{k_{\text{B}} T}{N_{\text{A}} d^3}  \left[ (1-\phi)\ln (1-\phi) + \chi \phi (1-\phi) \right] 
\end{equation}
Similarly for the elastic energy, we can compute the free energy density by dividing by the volume. 

We wish to minimize $\Delta F = \int \Delta f(\vb{x}) \dd \mathbf{x}$, where $\dd \mathbf{x}$ is the volume element. This shape optimization problem amounts to minimizing this free energy with respect to all the vertex positions $\vb{x}_i$ of the mesh, $\pdv*{\Delta F}{\vb{x}_i}=0$. To compute these derivatives, we use the chain rule, 
\begin{equation}
    \frac{\delta {F}}{\delta \mathbf{x}_i} = \frac{\delta {F}}{\delta N_\text{s}} \frac{\partial N_\text{s}}{\partial \mathbf{x}_i}
\end{equation}
and note that $\delta F/\delta N_\text{s} \propto -\Pi(\mathbf{x})$ as discussed in the previous section. Since each element's volume is given by $V = x N_\text{p} + N_\text{s}$, and $x$ and $N_\text{p}$ are constants, $\partial N_\text{s}/ \partial \mathbf{x}_i = \partial V/ \partial \mathbf{x}_i$, and hence we find,
\begin{equation}
    \frac{\delta {F}}{\delta \mathbf{x}_i} \propto -\Pi (\mathbf{x}) \frac{\partial V}{\partial \mathbf{x}_i}
\end{equation}
where $V(\mathbf{x})$ is the volume of the simplicial element and $\Pi(\mathbf{x})$ is the corresponding osmotic pressure. The volume is a known function of its vertices, and thus an analytical derivative of the free energy with respect to the shape of the hydrogel is obtained, facilitating high performance of the resulting code. We program this functional and its shape gradient in \textit{Morpho} \cite{Joshi2022b}. Within this environment, we can now minimize this functional in the presence of additional energies and constraints \cite{Joshi2022b} for arbitrary geometries in any dimension. Details of the \textit{Morpho} implementation are provided in the Appendix and codes are provided as Supplementary Information. 

In this work, we assume that the chains are uniformly distributed throughout the hydrogel, so $N_\text{c}$ does not depend on $\vb{x}$, but the formulation above and the implementation in \textit{Morpho} can be easily tweaked to allow a spatially varying initial $N_\text{c}$. It can be seen that we have three non-dimensional parameters, namely, the Flory-Huggins mixing parameter $\chi$, the relative strength of the elastic energy to the mixing energy $N_\text{c} \alpha^3/V_0$ and the reference volume fraction $\phi_0$ \cite{Quesada-Perez2011}. Given an initial value of $\phi$, we can vary these parameters to change the minima of the overall free energy. Thus, we can tune the volumetric swelling ratio, given by $r_\text{sw} = V_\text{f}/V_\text{i} = \phi_\text{i}/\phi_\text{f} = \phi_\text{i}/\phi_\text{eq}$, where the subscripts \emph{i} and \emph{f} refer to initial and final (equilibrium) states. Motivated by the experiments in Ref.~\cite{Louf2021}, we choose the values $\chi = 0.499$, $N_\text{c} \alpha^3/V_0=1$ and $\phi_0 \sim 0.036 $, which together set the equilibrium value $\phi_\text{eq} \sim 0.1$. Varying the initial volume fraction $\phi_\text{i}$ between $0.1$ and $1$, we get volumetric swelling ratios  $r_\text{sw} =\phi_\text{i}/\phi_\text{eq} \sim 1 - 10$, allowing us to access the range observed in the experiments.

We leverage a convenient hard constraint available in \textit{Morpho}, whereby vertices are excluded from a boundary defined by the contours or level-sets of a scalar function. Inspired by the experiments described in the introduction\cite{Louf2021}, we introduce hard-sphere beads surrounding the hydrogel. To mimic the experimental geometry, $N_\text{b}$ hard-spheres are distributed around the hydrogel sphere in contact with its surface. An illustrative example, Fig. \ref{fig:experiment}B, shows snapshots as the minimization proceeds from an initial spherical state for $N_\text{b}=30$, comparable to the experimental scenario, and depicts the final equilibrium state. To account for the projective imaging used in the experiment, we compute simulated images by rasterizing the configurations in 3D and summing them along a viewing axis as displayed in Fig. \ref{fig:experiment}C. 
\section{Results}

\begin{figure*}
\centering
\includegraphics{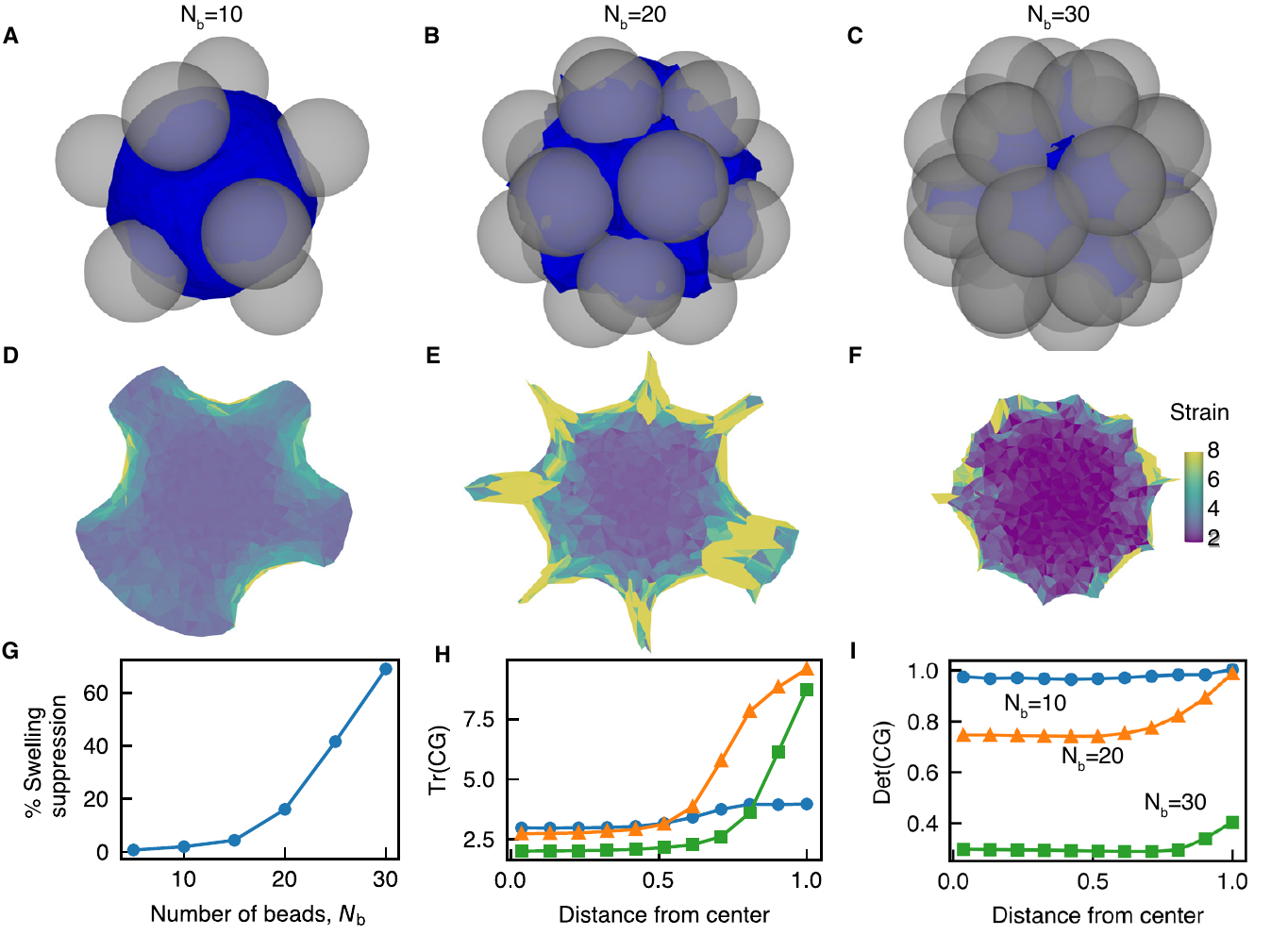}
\caption{Hydrogel Swelling in 3D. \textbf{A, B, C} Swollen hydrogels with $N_\text{b} = 10$, $20$ and $30$ surrounding beads respectively of radius $R_\text{b} = 0.7$. The swelling ratio is $r_\text{sw}=3$. \textbf{D, E, F} Trace of the Cauchy-Green tensor $\Tr(\mathbf{CG})$ sliced along the $x-y$ plane for the simulations in \textbf{A, B} and \textbf{C} respectively. \textbf{G} The \% suppression of swelling as a function of $N_\text{b}$. \textbf{H} $\Tr(\mathbf{CG})$, averaged along angular variables, as a function of distance from the center of the hydrogel, plotted for various number of beads. The distance is normalized by its maximum. \textbf{I} The corresponding plots for the determinant of the CG tensor.} 
\label{fig:cg}
\end{figure*}

We perform simulations of the swelling hydrogel for swelling ratios ranging from $r_\text{sw} \in [2, 6]$, inspired by experimental values, and with varying sizes ($R_\text{b}$)  and numbers ($N_\text{b}$) of confining beads.

In Fig.~\ref{fig:cg}A, B and C, we display the equilibrium configuration for a hydrogel sphere with 10, 20 and 30 adjacent beads respectively of $R_\text{b}=0.7$. 
We observe that the resulting configurations strongly resemble the morphologies observed in experiments as shown in Fig. \ref{fig:experiment}A and Ref. \cite{Louf2021}. We also display corresponding cross sections of these configurations with the state of strain in Fig.~\ref{fig:cg}D-F as will be discussed later. The final volume $V_\text{f}$ of the hydrogel is less than that of the final volume $V_\text{f,u}$ of the unconstrained hydrogel with the same parameters. We define the percent swelling suppression due to the confinement then as $c = (\Delta V_\text{f,u} - \Delta V_\text{f})/\Delta V_\text{f,u} \times 100$, where $\Delta V = V - V_\text{i}$ is the volume of the solvent absorbed. We plot this swelling suppression as a function of the number of beads $N_\text{b}$ in Fig.~\ref{fig:cg}G for a swelling ratio of $r_\text{sw}=3$ and bead radius $R_\text{b}=0.7$.
The increase in the swelling suppression with confinement is consistent with the observations in the experiments in Ref.~\cite{Louf2021}

Next, we reconstruct the state of strain in the deformed gel as follows. First, we run a corresponding simulation without the bead constraints to obtain the unconfined swollen profiles. By comparing the elements and their vertex positions in the constrained and unconstrained swollen meshes, we can compute the Cauchy-Green strain tensor ($\mathbf{CG}$) for each element as follows. First, we compute a Gram matrix for every element $V_k$ in the confined mesh,
\begin{equation}
    G^k_{ij} = \mathbf{s}_i \cdot \mathbf{s}_j, \quad i \in \{1,2,3\}
\end{equation}
where, $\vec{s_i} = \vec{v_i}^k-\vec{v_0}^k$ is the vector connecting the $0^{th}$ and $i^{th}$ vertex of the element. We also compute the corresponding Gram matrix for the reference unconfined element, 
\begin{equation}
    (G^k_\text{ref})_{ij} = \mathbf{s}^r_i \cdot \mathbf{s}^r_j, \quad i \in \{1,2,3\}.
\end{equation}
From these quantities, we compute the Cauchy-Green tensor for the volume element $V_k$ as,
\begin{equation}
    CG^k_{ij} = (G^k_\text{ref})^{-1}_{il} \ G^k_{lj}, \quad i \in \{1,2,3\}.
\end{equation}
In the present work, both constrained and unconstrained meshes have the same topology. It is however important that if refinement or element exchanges are performed during optimization that they be executed on both meshes to preserve an element-to-element map between the two final states for the $\mathbf{CG}$ tensor calculation to be valid. 

In Fig.~\ref{fig:cg}H, we plot the Trace of this tensor ($I_1 = \Tr(\mathbf{CG})$), averaged over angular variables, as a function of the distance from the center, thus probing the strain due to confinement. We also display similar plots of the determinant of this tensor ($I_3 = \det (\mathbf{CG})$), which corresponds to the local volume change, in Fig.~\ref{fig:cg}I. We can reconstruct and visualize these quantities spatially. Example profiles of $\Tr(\mathbf{CG})$ sliced across the $x-y$ plane for the simulations in Fig.~\ref{fig:cg}A, B and C are shown in Fig.~\ref{fig:cg}D, E and F respectively.

From this analysis, we observe that the strain is largely confined to the outer half of the hydrogel sphere, and is increasingly localized to the outer extremities as we increase the number of beads. For a small number of beads, the strain profiles are markedly different from those expected from a uniform spherical confinement, which would result in a constant value of $\Tr(\mathbf{CG})$ throughout the hydrogel. Indeed, the localization of strain \textit{a posteriori} justifies a key assumption of the model developed in Ref.~\cite{Louf2021}: that the sphere can be decomposed into an undeformed core and a highly deformed outer region. 
\begin{figure*}
\centering
\includegraphics{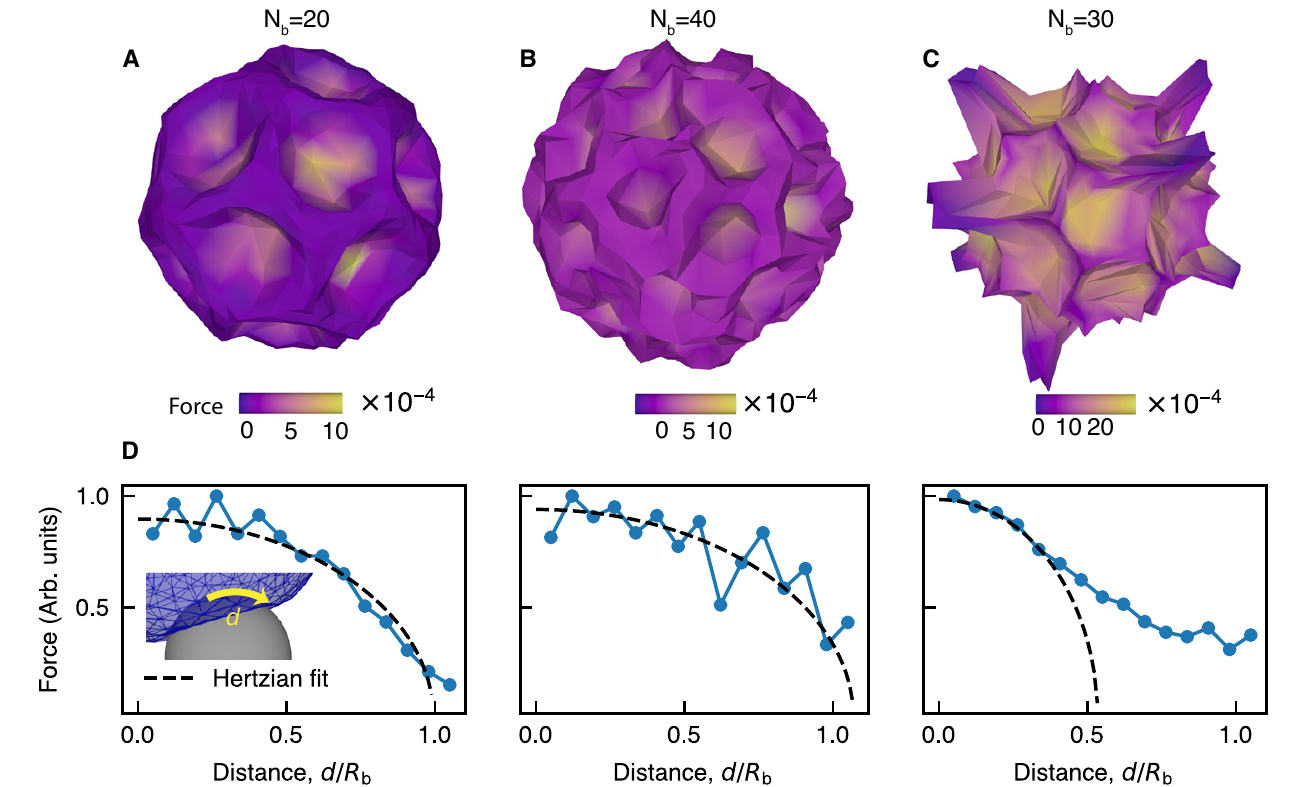}
\caption{Contact forces for a swollen confined hydrogel. \textbf{A, B, C} Gradient of the hydrogel functional at the boundary of the hydrogel for number of beads $N_\text{b}$,  swelling ratio $r_s$  and bead radius $R_\text{b}$ equal to $(20, 2.4, 0.5)$, $(40, 3.0, 0.3)$ and $(30, 2.4, 0.7)$ respectively. \textbf{D} The average contact force as a function of the normalized arc-length $d/R_\text{b}$ away from the center of the contact (as illustrated in the inset) for the simulations in A, B and C. The dashed line shows a fit to the Hertzian contact pressure profile, $p=a \sqrt{1-(d/d_0)^2)}$. We can see that the Hertzian model works well when the contacts are independent, but starts to diverge away from it at larger $d/R_\text{b}$ values when the contacts overlap.}
\label{fig:contact}
\end{figure*}

To understand the contact mechanics of the gel-constraint interface, we compute the contact forces between the hydrogel and the beads as follows. We obtain the swelling force at each mesh vertex by taking the gradient of the free energy (Eq.~\ref{eq:hg}) with respect to the vertex positions. A few examples of this force at the boundary are plotted in Fig.~\ref{fig:contact}A, B and C. As expected, this force is largest at the center of the contacts due to the constraints. To probe the mechanics of the hydrogel-bead contact, we plot the 1D profiles of the contact pressure, averaged over 1D boundary slices of the forces and over all the beads. As shown in Fig.~\ref{fig:contact}D, this profile follows the Hertzian contact mechanics estimate of $p=a \sqrt{1-(d/d_0)^2)}$ close to the contact point. This provides further support for the applicability of Hertzian contact mechanics assumed in the model developed in Ref.~ \cite{Louf2021}. We also see in the right panel of Fig.~\ref{fig:contact}D that the extent to which a Hertzian profile is followed decreases when the contacts overlap, echoing the analysis in Ref.~ \cite{Louf2021} for stronger confinements.
\section{Discussion}
In this paper, we formulated a general approach to determining the equilibrium configurations and properties of swollen hydrogels under arbitrary confinement in 3D. Recasting the state of the system as the solution to an energy optimization problem enables us to take advantage of optimization theory and permits convenient enforcement of constraints. 

We applied this framework to understand the mechanics of hydrogel beads confined between rigid spherical beads as has been studied experimentally. Our resulting numerically optimized configurations give fresh insight into the unusual mechanical properties of these gels that are not experimentally accessible. Notably, we observe a transition in the behavior as a function of confinement: for small enough confinements, the strain is localised on the outside of the sphere with the core of the hydrogel remaining relatively strain-free. At higher confinement, the strain becomes more and more uniformly distributed throughout the hydrogel matching the configuration expected for a sphere swelling with a fixed outer boundary. We also examined the contact mechanics of the gel-bead system, finding a transition from Hertzian to non-Hertzian behavior in the contact pressure distribution as a function of swelling. 

While we investigated homogeneous hydrogels, our framework can be easily used to impose a position-dependent swelling ratio, elasticity, etc. which can be used in applications such as hydrogel bi-layers \cite{Cangialosi2017,Schaffter2022,Bayles2022,Pantula2022} or other functional hydrogels. Further, it can be readily extended to include ionic contributions \cite{Quesada-Perez2011}.

We formulated and solved the problem using our open source shape optimization environment \textit{Morpho}, which means that we can readily accommodate a number of experimentally relevant extensions to the model. We could easily incorporate other energies such as gravitational potentials, electric fields, surface tension and surface elasticity, for example.  By allowing the constraining beads to move with a pinning energy, we could model situations where the confinement pressure is finite. Our method could also be adapted to study the kinetics of swelling in the quasistatic limit where experiments typically take place by recasting the optimization problem as a gradient-flow problem with an appropriate time-stepping scheme. In addition, careful application of refinement could be used to accommodate topological changes, or strain dependent connectivity energies could be incorporated in order to study fracture of hydrogels \cite{Yang2018,Lin2019}.

\section*{Author Contributions}
CJ, MQG and TJA developed the theoretical model and implemented the code. CJ obtained the simulation data and analyzed the results. J-FL and SSD designed the experiment and obtained the experimental images. All authors contributed to preparing the manuscript. 

\section*{Conflicts of interest}
There are no conflicts to declare.

\section*{Appendix}

\subsection*{Osmotic Pressure}
Conventional presentations of hydrogel swelling rely on a pressure-balance approach. In this appendix, we show that our optimization formalism reduces to the regular theory. 
To do so, begin by noting that the mixing contribution to the free energy, $\Delta F_{\text{mix}}$, is given by the Flory-Huggins theory. The osmotic pressure contribution from this energy is 
\begin{equation}
    \Pi_{\text{mix}} = -\frac{N_{\text{A}}}{\nu_\text{s}} \pdv{\Delta F_{\text{mix}}}{N_\text{s}}
\end{equation}
Because the volume fraction $\phi$ depends on $N_\text{s}$,
\begin{equation}
    \phi = \frac{x N_\text{p} }{(xN_\text{p} + N_\text{s})},
    \label{eq:phi}
\end{equation}
we may re-express derivatives with respect to $N_\text{s}$ using the chain rule,
\begin{align*}
    \pdv{N_\text{s}} &= \pdv{\phi}{N_\text{s}} \pdv{\phi} \\
    &= - \frac{xN_\text{p}}{(xN_\text{p} + N_\text{s})^2} \pdv{\phi} \\
    &= -\frac{1}{x N_\text{p} } \phi^2 \pdv{\phi}.
\end{align*}
Hence,
\begin{align*}
    \frac{1}{k_{\text{B}} T}&\pdv{\Delta F_{\text{mix}}}{N_\text{s}} = \pdv{N_\text{s}} \{ N_\text{s} \left[ \ln (1-\phi) + \chi \phi \right] \} \\
    &= \left[ \ln (1-\phi) + \chi \phi \right]  \\
    &\quad + N_\text{s} \left( -\frac{1}{xN_\text{p}} \phi^2  \right) \pdv{\phi} \left[ \ln (1-\phi) + \chi \phi \right] \\
     &= \left[ \ln (1-\phi) + \chi \phi \right] 
    - \left( \frac{N_\text{s}}{xN_\text{p}} \phi^2  \right) \left[ \frac{-1}{(1-\phi)} + \chi  \right]
\end{align*}
We rearrange  Eq.~\eqref{eq:phi}, 
\begin{equation*}
    \frac{N_\text{s}}{x N_\text{p} } = \frac{1}{\phi}-1 = \frac{(1-\phi)}{\phi},
\end{equation*}
and use this to eliminate $N_\text{s}/x N_\text{p}$ from the osmotic pressure,
\begin{align*}
    \frac{1}{k_{\text{B}} T}\pdv{\Delta F_{\text{mix}}}{N_\text{s}} &= \left[ \ln (1-\phi) + \chi \phi \right] 
    - \left( \frac{N_\text{s}}{x N_\text{p}} \phi^2  \right) \left[ \frac{-1}{(1-\phi)} + \chi  \right] \\
    &= \left[ \ln (1-\phi) + \chi \phi \right] 
    - (  \phi (1-\phi) ) \left[ \frac{-1}{(1-\phi)} + \chi  \right] \\
    &= \left[ \ln (1-\phi) + \chi \phi \right] + \phi - \chi \phi (1-\phi) \\
    &= \phi + \ln (1-\phi) + \chi \phi^2.
\end{align*}
We hence recover the standard result, expressed for example as Eq. (8) from Ref.~\cite{Quesada-Perez2011}:
\begin{equation}
    \Pi_{\text{mix}} = -\frac{N_{\text{A}}}{\nu_\text{s}} \pdv{\Delta F_{\text{mix}}}{N_\text{s}} = -\frac{N_{\text{A}} k_{\text{B}} T}{\nu_\text{s}} \left[ \phi + \ln (1-\phi) + \chi \phi^2 \right]
\end{equation}

Note that in the literature, this osmotic pressure is sometimes expressed in terms of an `effective diameter' of the solvent molecule \cite{Louf2021}:
\begin{equation}
    \Pi_{\text{mix}} = -\frac{k_{\text{B}} T}{d^3} \left[ \phi + \ln (1-\phi) + \chi \phi^2 \right],
\end{equation}
which can be readily understood, since it implies $d^3 = \nu_\text{s} / N_{\text{A}}$. 

\subsection*{Simulation details}
To compute the structure of the hydrogel in \textit{Morpho}, we start by constructing an initially spherical \textbf{Mesh} corresponding to the unit ball $\left|\mathbf{x}\right|^{2}<1$ with Morpho's \textit{meshgen} module. An \textbf{OptimizationProblem} object is then defined and a \textbf{Hydrogel} functional, implementing the above discussed free energy density, is added to it. For hard confinements, we define level-set constraints corresponding to the objects (spheres, ellipsoids, planes, etc.) through the \textbf{ScalarPotential} object from the \textit{functionals} module. A \textbf{ShapeOptimizer} object is then created to optimize the shape. We perform gradient descent with a fixed step size. A \textbf{Volume} object is used to keep track of the volume of the hydrogel during relaxation. 

To initialize the positions of the hard spheres, we define a dummy shell mesh with radius $R + R_\text{b}$ with $N_\text{b}$ number of vertices placed randomly. We first confine the vertices to lie on the shell by using a \textbf{ScalarPotential} object. We then define an electrostatic repulsive pairwise interaction between the vertices using a \textbf{PairwisePotential} object from the \textit{functionals} module, thus proceeding to solve the Thomson problem. The resulting mesh vertex positions are used as the sphere centers for the level set constraints. We thus get equidistantly packed spheres on the outer shell. 

All 3D visualizations are made using the \textit{povray} module. The slices of the Cauchy Green strain tensor's trace are generated using the \textit{meshslice} module.
\begin{table}
\centering
\begin{tabular}{|c|c||c|} 
 \hline
 Parameter & Symbol & Value\\ 
 \hline
 \hline
 Flory-Huggins parameter & $\chi$ & 0.499  \\ 
 \hline
 Relative elastic strength & $N_\text{c} \alpha^3/V_0$ & 1 \\
 \hline
 Reference volume fraction & $\phi_0$ & 0.036 \\
 \hline
 Confining bead radius & $R_\text{b}$ & [0.5, 1] \\
 \hline 
 Swelling ratio & $r_\text{sw}$ & [2,6] \\
 \hline
 Number of beads & $N_\text{b}$ & [5,10,\ldots 40] \\
 \hline
\end{tabular} 
 \caption{Parameters used for the hydrogel swelling simulations.\label{table:parameters}}
\end{table}

The parameters used in the simulations are listed in Table \ref{table:parameters}.

\section*{Acknowledgements}
The authors thank Abigail Plummer for useful discussions. This material is based upon work supported by the National Science Foundation under Grant No. ACI-2003820 (CJ, MQG and TJA) and Grant No. DMR-2011750 (JFL and SSD). This material is also based upon work by SSD supported by the U.S. Department of Energy's Office of Energy Efficiency and Renewable Energy (EERE) under the Geothermal Technologies Office (GTO) INNOVATIVE METHODS TO CONTROL HYDRAULIC PROPERTIES OF ENHANCED GEOTHERMAL SYSTEMS Award Number DE-EE0009790.

\balance

\bibliographystyle{rsc} 
\providecommand*{\mcitethebibliography}{\thebibliography}
\csname @ifundefined\endcsname{endmcitethebibliography}
{\let\endmcitethebibliography\endthebibliography}{}

\end{document}